\documentclass[preprint, showpacs]{revtex4-1}
\usepackage{hyperref}

\usepackage{amsfonts}
\usepackage{amsmath}
\usepackage{amssymb}
\usepackage{graphicx}
\newtheorem{thm}{Theorem}
\newtheorem{cor}{Corollary}
\newtheorem{lemma}{Lemma}

\newtheorem{prop}{Proposition}

\def\qed{\hfill $\blacksquare$}

\def\tr{\hbox{Tr}}

\def\Hil{{\cal H}}
\def\C{\mathbb{C}}

\def\tensor[#1]{ \C^{#1} \ot \C^{#1}}
\def\be{\begin{eqnarray}}
\def\ee{\end{eqnarray}}
\def\bee{\begin{eqnarray*}}
\def\eee{\end{eqnarray*}}

\def\ts{\textstyle}
\def\bra{\langle}
\def\ket{\rangle}
\def\kb{ \ket \bra }

\def\rt2{\ts \frac{1}{\sqrt{2}} }

\def\ot{\otimes}

\begin{document}

\title{Three maximally entangled states can require two-way local operations and classical communications for local discrimination}
\author{Michael Nathanson} 
\affiliation{Department of Mathematics and Computer Science, Saint Mary's College
of California, Moraga, CA, 94556, USA}
\email{man6@stmarys-ca.edu}
\pacs{03.67.Mn, 03.67.Hk}

\begin{abstract}
We show that there exist sets of three mutually orthogonal $d$-dimensional maximally entangled states which cannot be perfectly distinguished using one-way local operations and classical communication (LOCC) for arbitrarily large values of $d$. This contrasts with several well-known families of maximally entangled states, for which any three states can be perfectly distinguished. We then show that two-way LOCC is sufficient to distinguish these examples. We also show that any three mutually orthogonal $d$-dimensional maximally entangled states {\it can} be perfectly distinguished using measurements with a positive partial transpose (PPT) and can be distinguished with one-way LOCC with high probability. These results circle around the question of whether there exist three maximally entangled states which cannot be distinguished using the full power of LOCC; we discuss possible approaches to answer this question.
\end{abstract}

\maketitle
\section{Introduction}
Although much progress has been made in recent years, entanglement remains one of the most mysterious phenomena in the quantum world, and the interplay between entanglement and locality is the engine behind many quantum protocols. In the paradigm of local bipartite quantum state discrimination, two parties (Alice and Bob) each control a quantum system, which we represent as finite-dimensional spaces $\Hil_A$ and $\Hil_B$. Their joint system $\Hil_A \ot \Hil_B$ has been prepared in a pure state from the set $S = \{\vert \psi_i \ket\}_{ i = 0, \ldots N-1}$; the two component systems are then separated. Alice and Bob know $S$ and they would like to determine the value of $i$. Since they are physically separate, their possible measurement protocols are restricted to those using only local quantum operations and classical communication (LOCC). The study of the power and limitations of LOCC has potential applications in cryptography, communication, and data hiding \cite{data-hiding, MatthewsWehnerWinter} and is also of inherent interest as a tool to understand entanglement. 

\begin{figure}\label{LOCC Subsets}
\includegraphics{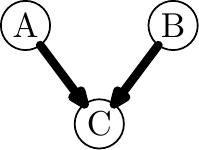} \hfil \includegraphics{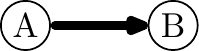} \hfil \includegraphics{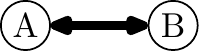}\caption{Schematic for (a) local operations (b) one-way LOCC (c) full LOCC}
\end{figure}

The class of bipartite LOCC measurements can be further broken down based on how the classical communication is used. Local product measures are those in which Alice and Bob separately perform measurements and only communicate after the fact to compare and interpret their results. In one-way LOCC, Bob may adapt his measurement based on classical information received from Alice but no information is allowed to move in the other direction. Finally, full two-way LOCC allows Alice and Bob to communicate classically as much as they like and to iteratively adapt their measurements as they go. These distinctions are depicted schematically in Figure \ref{LOCC Subsets}.

In this paper, we will be exclusively concerned with families of orthogonal maximally entangled states. Orthogonal states could be perfectly distinguished if both parties were in the same place, while sets of states with maximal entanglement have the property that neither party can unilaterally extract any information about the identity of $\vert \psi_i\ket$; both parties must make measurements in order to learn anything at all. We will be primarily concerned with the question of {\it perfect} local state discrimination: When is it possible for Alice and Bob to determine the identity of $\vert \psi_i\ket$ 100\% of the time? 

It is known that any two orthogonal maximally entangled states may be perfectly distinguished with one-way LOCC (since any two pure states can be  distinguished this way \cite{Walgate, Virmani-2001}). In \cite{Nathanson2005}, we showed that any three mutually orthogonal maximally entangled $\mathbb{C}^3 \otimes \mathbb{C}^3$ states can be distinguished with local operators and classical communications, and Fan \cite{Fan} showed that any three generalized Pauli states in dimension $d$ can be perfectly distinguished if $d$ is a prime number greater than two. Since then it has been an open question whether {\it any} three mutually orthogonal maximally entangled states in high dimensions can be distinguished perfectly with LOCC. 

In this work we give examples of three maximally entangled non-qubit states which cannot be perfectly distinguished with one-way LOCC. These examples {\it can} be distinguished using full two-way LOCC. These are obviously the smallest sets of states with this property and may be the first known sets of entangled pure states for which two-way LOCC is necessary and sufficient.  

We also show that any three orthogonal maximally entangled states may be distinguished with positive partial transpose measurements (PPT) and can be distinguished with one-way LOCC {\em with high probability}. The question of distinguishing any three maximally entangled states using full LOCC measurements remains open, but these results show that the answer lies in the space between PPT and one-way LOCC measures and that this space is small.

The rest of the paper is organized as follows: In Section \ref{triples}, we present a summary of the results which pertain to local discrimination of a generic triple of orthogonal maximally entangled states. In Section \ref{CounterExamples}, we discuss the necessary conditions for one-way LOCC discrimination and construct two different families of examples for which one-way LOCC is insufficient. We also point out that it {\it is} possible to distinguish these families using LOCC with two rounds of communication. In Section \ref{PPTSection} we prove our PPT bound, while Section \ref{OneWayConstruction} proves an upper bound on the minimum error in one-way LOCC. Section \ref{MoreStates} extends these results to larger sets of states which are distinguishable with PPT but not one-way LOCC. Finally, we conclude with a discussion of the question which is conspicuously not answered here--whether there exist triples of orthogonal maximally entangled states which cannot be locally distinguished. There is also an appendix, in which we present two-way protocols to distinguish the example sets from each other. 

\section{Summary of results for three maximally entangled states}\label{triples}
In the interest of studying locality and entanglement, we would very much like to understand which sets of bipartite states can and cannot be distinguished using LOCC. However, the class of LOCC measurements is notoriously messy to describe mathematically, so we approach the problem using the standard nested set of measurement classes: 
\be \label{hierarchy}
LO \subset LOCC-1 \subset LOCC \subset SEP \subset PPT \subset ALL
\ee
Here, LOCC represents the full class of measurements which can implemented with LOCC. The subsets of LOCC are local product measures (LO) and one-way LOCC (LOCC-1), which have already been discussed. The supersets of LOCC are described in terms of mathematical formalism, not operationally. A measurement $\mathbb{M} = \{M_k\}$ is separable (SEP) if each $M_k$ can be decomposed as $M_k = \sum_l A_{k,l} \ot B_{k,l}$ across the $A:B$ split, with $A_{k,l} \ot B_{k,l} \ge 0$ for each $k,l$. $\mathbb{M}$ is a positive partial transpose (PPT) measurement if, for each $M_k$, the operator $({\cal I} \ot {\cal T})M_k$ is positive semidefinite, where ${\cal T}$ is the transpose operation. This is described more in Section \ref{PPTSection}. 

The inclusions in (\ref{hierarchy}) are strict, and for each separation we can find sets of states which demonstrate this separation. Table \ref{TableResults} summarizes the state of knowledge about small sets of maximally entangled states, highlighting one or two citations for each. In the table, $k$ is the number of states and $d$ is the dimension; and the question is: Given a set of $k$ orthogonal maximally entangled states in $\tensor[d]$, are you able to distinguish them perfectly? 
\begin{table}\centering
\begin{tabular}%{r  c ccc c }
{|r | c| c|c|c| c |}
\hline\hline
& PPT & LOCC & LOCC-1 & LO& Citation \cr\hline
$k = 2$, $d=2$ &Always & $\longleftarrow $& $\longleftarrow$ & Always& \cite{Walgate} \cr \hline 
$k = 3$, $d=2$ & Never & $\longrightarrow $& $\longrightarrow$ & Never & \cite{Walgate-2002} \cr \hline 
$k = 3$, $d=3$ & Always & $\longleftarrow $& $\longleftarrow$ & Always & \cite{Nathanson2005} \cr \hline 
$k = 3$, $d=4$ & {\bf Always} & {\bf ???????}  & {\bf Sometimes} & Sometimes &   \cr \hline 
$3$ Pauli states &Always & $\longleftarrow $& $\longleftarrow$ & Always& \cite{Fan} \cr  $d \ge 3$ prime& &&&&
\cr \hline 
$k = 4$, $d=4$ & Sometimes &Sometimes  &Sometimes & Sometimes & \cite{BGK,Cosentino, FourQuquad, OH-2006} \cr \hline 
$k > d \ge 2$ & Never & $\longrightarrow $& $\longrightarrow$ & Never & \cite{ Nathanson2005, Cosentino,Ghosh-2001,Hayashi Bounds 2006} \cr \hline  \hline
\end{tabular}\caption{Known results for distinguishing maximally entangled states}\label{TableResults} \end{table}

The fourth line ($k = 3, d = 4$) and its associated results are contained in the current paper. 

Note that the standard results for three Pauli states, for three maximally entangled states in $\tensor[3]$, and for two qubit maximally entangled states are generally stated in terms of one-way LOCC. Here we simply point out they can actually be accomplished with a product measurement. The proof of Fan's result \cite{Fan} given in \cite{Nathanson2005} shows that Alice can measure in a basis of eigenstates for a generalized Pauli matrix. Bob will always measure in the same basis as Alice and does not need to know her outcome to do so. They can then separately send their results to a third party who will identify the state they started with, as indicated in Figure \ref{LOCC Subsets}. (In the special case that three matrices generate a cyclic group, the product measurement in \cite{OH-2006} may also be used.) Similarly, the measurement constructed in \cite{Nathanson2005} to distinguish maximally entangled states in $\tensor[3]$ is actually a product measurement, although not explicitly stated. Finally, we point out that even the standard result for two pure states becomes a product measurement when both states are qubit maximally entangled states.

As is apparent from the table, we do not yet  know whether all triples of orthogonal maximally entangled states can be distinguished with LOCC. The focus of the current work is to explore this question. The first result is a negative one: 
\begin{thm} \label{LOCC1} 
There exist triples of mutually orthogonal maximally entangled states in $\mathbb{C}^4 \otimes \mathbb{C}^4$ which cannot be distinguished with one-way LOCC. 

In fact,  there exist triples of mutually orthogonal maximally entangled states in $\mathbb{C}^d \otimes \mathbb{C}^d$ which cannot be distinguished with one-way LOCC in any dimension $d$ for which $d$ is even or $d \equiv 2  \mbox{ mod } 3$. 
\end{thm}
These examples are significant, as there are not many known sets which highlight the difference between one-way LOCC and full LOCC. The nonlocality without entanglement states of Bennett {\it et al.}\cite{Bennett-I-99} are a basis of nine pure product states in $\tensor[3]$ which cannot be distinguished with LOCC--they exemplify the difference between separable measurements and LOCC. In addition, a subset of seven of these nine states can be distinguished with full LOCC but not one-way LOCC. This is the most famous example of pure states which have this property. In fact, few others have been found. Bandyopadhyay {\it et al.}\cite{BGK} gave examples of sets of four and five generalized Pauli states which cannot be distinguished with one-way LOCC; but it remains an open question whether they can be distinguished with LOCC at all. Hayashi {\it et al.}\cite{OwariHayashi} showed that one can distinguish a pure maximally entangled states from the maximally noisy mixed state  more successfully with two-way LOCC than with one-way; but the two states are not perfectly distinguishable in either case. Thus, the examples in Section \ref{CounterExamples} are interesting in their own right, to understand the difference between these two paradigms. 

While LOCC-1 is a strict subset of all possible LOCC measurements, the positive partial transpose (PPT) measurements form a strict superset. And in this case, we get an affirmative result: 
\begin{thm} \label{PPT}
Any three orthogonal maximally entangled states in $\mathbb{C}^d \ot \mathbb{C}^d$, $d \ge 3$, can be perfectly distinguished with a PPT measurement. 
\end{thm}
When $d = 3$, this theorem follows from the LOCC result in \cite{Nathanson2005}. For larger values of $d$, it is a corollary to a more general result, Theorem \ref{PPTTheorem}, which is stated and proved in Section \ref{PPTSection}. 

Taken together, Theorems \ref{LOCC1} and \ref{PPT} show that there is a gap between LOCC-1 and PPT when it comes to triples of maximally entangled states. Since the set of LOCC measurements is strictly between these two sets, we would like to understand how big is the gap between them. The following result says that this gap is not very big. 
\begin{thm}\label{LOCC1 Bound}
Any three orthogonal maximally entangled states in $\mathbb{C}^d \ot \mathbb{C}^d$ can be distinguished with one-way LOCC with error probability at most $\frac{2}{3d}$. 
\end{thm} 
The question of whether any three orthogonal maximally entangled states can be distinguished using LOCC measures has proved challenging, as the answer lies in the small space between LOCC-1 and PPT. The significance of the current work is to show that there is in fact a gap between these two sets with respect to this problem; this will hopefully point the way to answering the general LOCC question. 

In the remainder of the paper, we prove our results and explore our examples in greater detail. 
\section{Sets which cannot be distinguished with one-way LOCC}\label{CounterExamples}
In this, section we present our families of examples. The first family is defined in $\tensor[d]$ when $d$ is even; the second family when $d \equiv 2 \mbox{ mod } 5$. First, we explicitly give necessary and sufficient conditions for one-way LOCC state discrimination. 
\subsection{Characterizing one-way LOCC measurement}
A one-way LOCC measurement is of the form $\mathbb{M} = \{ A_k \ot B_{k,j} \}$ with $\sum_k A_k = {\cal I}_A$ and $\sum_j B_{k,j} = {\cal I}_B$ for each value of $k$. If we get the outcome $A_k \ot B_{k,j}$, we conclude that our state was prepared as $\vert \psi_j \ket$. This will distinguish our states perfectly if, for each $k$ and $i \ne j$.
\bee
\bra \psi_i \vert A_k \ot B_{k,j} \vert \psi_i \ket = 0 \eee
This implies that if $\vert a \ot b \ket$ is an eigenvector of $A_k \ot B_{k,j}$ with nonzero eigenvalue, 
\bee
\bra \psi_i \vert a \ot b \ket \bra a   \ot b \vert \psi_i \ket = \left \vert \bra \psi_i \vert a \ot b \ket \right\vert^2 = 0
\eee
Thus without loss of generality, we can relabel and assume that $A_k = m_k \vert \overline{a_k} \kb \overline{a_k} \vert$ is a rank one matrix with trace $m_k$. Here, $\vert \overline{a_k}\ket$ is the entrywise complex conjugate of $\vert {a_k}\ket$ in the standard basis. Using the representation $\vert \psi_i \ket = (I \ot U_i)\vert \Phi \ket$, the (non-normalized) state of Bob's  system after Alice's measurement is $U_i\vert a_k \ket$. These can be distinguished if and only if the states $\{ U_i \vert a_k \ket\}_{i = 0\ldots N-1}$  form an orthonormal set.  This gives us our result: 
\begin{prop}\label{OneWayLOCC}
Given a set of states $S = \{ \vert \psi_i \ket = (I \ot U_i)\vert \Phi\ket \} \subset \mathbb{C}^d \ot \mathbb{C}^d$, with $\vert \Phi \ket$ the standard maximally entangled state. 

The elements of $S$ can be perfectly distinguished with one-way LOCC if and only if there exists a set of states $\{ \vert \phi_k \ket \} \subset \C^d$ and a set of positive numbers $\{ m_k \}$ such that  $\sum_k m_k \vert \phi_k \kb \phi_k \vert = {\cal I}_d$ and 
\be \label{Eqn One Way}
\bra \phi_k \vert U_j^*U_i \vert \phi_k \ket = 0
\ee
whenever $i \ne j$.

Equivalently, the elements of $S$ can be perfectly distinguished with one-way LOCC if and only if there exists a $d \times r$ partial isometry $W$ such that $WW^* = {\cal I}_d$ and such that whenever $i \ne j$, the $r \times r$ matrix $W^*U^*_iU_jW$ has every diagonal entry equal to zero.
\end{prop}
In the case of maximally entangled states, the $U_i$ are unitary matrices, which implies that $\bra \phi_k \vert U_j^*U_i \vert \phi_k \ket = \delta_{i,j}$ for each $k$. This also means that Alice's initial measurement provides no information about the identity of the prepared state. 

Bandyopadhyay, et al., studied maximally entangled states and one-way LOCC in \cite{BGK}, providing examples of four orthogonal maximally entangled states which cannot be distinguished with one-way LOCC. Their Corollary 3 gives a necessary condition for one-way LOCC discrimination, that there exists at least one vector $\vert \phi_k \ket$ which satisfies (\ref{Eqn One Way}). Proposition \ref{OneWayLOCC} gives a stronger necessary condition which is also sufficient; our examples will need this stronger result in cases when $d>4$.

\subsection{Counterexample in dimension $d=2m$} \label{CounterexamplesEven}
It is well known that no three maximally entangled states in dimension $d = 2$ can be perfectly distinguished with LOCC; we will use this fact to construct a counterexample in higher dimensions. Recall the qubit Pauli matrices :
\bee
X = \begin{pmatrix} 0 & 1 \\ 1 & 0 \end{pmatrix}, \qquad Y =  \begin{pmatrix} 0 & -i \\ i& 0 \end{pmatrix}, \qquad Z=  \begin{pmatrix}1 &0 \\ 0 & -1 \end{pmatrix}. \eee
These correspond to the Bell states $\{ \vert \Phi_1 \ket, \vert \Phi_2 \ket, \vert \Phi_3\ket \} \subset \tensor[2]$. 
We will use the Pauli matrices to build maximally entangled states in any even dimension. We will also be multiplying by an arbitrary phase. For any $\omega$ with $|\omega| = 1$, we define $T_\omega$ as the $m \times m$ diagonal unitary matrix with all ones except for the first entry: 
\bee
T_\omega:= \begin{pmatrix}\omega&&& \\ & 1&& \\ &&\ddots & \\ &&&1 \end{pmatrix}.
\eee
We then choose $\omega$ and $\gamma$ generically on the unit circle, so that none of $\omega, \gamma,$ or $\overline{\omega}\gamma$ is real.

Let $\vert \psi_0 \ket$ be the standard maximally entangled state. Setting $U = T_\omega \ot X$ and $V = T_\gamma \ot Z$, we can look at the maximally entangled states given by $S = \{ \vert \psi_0 \ket, (I \ot U) \vert \psi_0 \ket, (I \ot V) \vert \psi_0 \ket\} \subseteq \mathbb{C}^{2m}\ot \mathbb{C}^{2m}$. It is easily checked that these are mutually orthogonal and maximally entangled. 

In the case $d = 4$, the three states are given by
\be
\vert \psi_0 \ket &=& \frac{1}{2} \left( |00\ket + |11\ket\right)_{A_1B_1} \ot \left( |00\ket + |11\ket\right)_{A_2B_2} \nonumber \\
\vert \psi_1 \ket &=& \frac{1}{2} \left( \omega |00\ket + |11\ket\right)_{A_1B_1} \ot \left( |01\ket + |10\ket\right)_{A_2B_2}  \label{d4 example} \\
\vert \psi_2 \ket &=& \frac{1}{2} \left( \gamma |00\ket + |11\ket\right)_{A_1B_1} \ot \left( |00\ket - |11\ket\right)_{A_2B_2}  \nonumber
\ee
We now show that these examples cannot be distinguished with one-way LOCC. 

Suppose Alice performs an initial measurement $\mathbb{M}$ on her system and receives the outcome corresponding to the operator $M^T$. Note that the outcome yields no classical information about the identity of the state (since the states are maximally entangled) and that after measurement the system is in the (nonnormalized) state $(I \ot U_iM^{1/2})\vert \psi_0\ket$. In order for perfect discrimination to happen, we need $\tr U_iMU_j^*= 0$ whenever $i \ne j$. We write $U,V,$ and $M$ as a block matrices, in blocks of size $2$ and $2(r-1)$, where ${\cal I}$ is the $(r-1) \times (r-1)$ identity matrix: 
\bee
U = \begin{pmatrix} \omega X & \\ & X \ot {\cal I} \end{pmatrix}, \qquad V = \begin{pmatrix}\gamma Z &  \\ & Z \ot {\cal I} \end{pmatrix}, \qquad M =  \begin{pmatrix}A & C^* \\ C & B \end{pmatrix} \ge 0.
\eee
The required orthogonality conditions imply that \be
d \bra \psi_0 \vert M \vert \psi_1 \ket &=& \tr UM= \omega \tr XA + \tr(X \ot {\cal I})B = 0 \nonumber \\
d \bra \psi_0 \vert M \vert \psi_2 \ket &=&\tr VM= \gamma \tr ZA +\tr(Z \ot {\cal I})B = 0 \label{EvenOrthConditions} \\
d \bra \psi_1 \vert M \vert \psi_2 \ket &=&\tr VMU^*= -i\overline{\omega}\gamma \tr YA  - i \tr \tr(Y \ot {\cal I})B = 0 \nonumber
\ee
Since $A,B,X,Y,Z$ are all hermitian and the product of two hermitian matrices always has a real-valued trace, we see that \bee
\omega \tr XA & =&-\tr(X \ot {\cal I})B  \in \mathbb{R} \Longrightarrow \tr XA = 0, \\
\gamma \tr ZA & =&-\tr(Z \ot {\cal I})B  \in \mathbb{R} \Longrightarrow \tr ZA = 0, \\
\overline{\omega}\gamma \tr YA & =&-\tr(Y \ot {\cal I})B  \in \mathbb{R} \Longrightarrow \tr YA = 0.
\eee
Since the Pauli matrices form a basis for $2\times 2$ hermitian matrices, we are forced to conclude that $A = t {\cal I}_2$ for some $t \in [0,1]$.  

From Proposition \ref{OneWayLOCC}, to distinguish these states with one-way LOCC, we need Alice to have complete measurement $\mathbb{M} = \{ M_i \}$ consisting of rank one matrices. If $A$ is a multiple of the identity matrix, then either $A = 0$ or else the rank of $M$ is at least two. Thus, either $\mathbb{M}$ contains measurements of rank greater than one or else $\sum_i M_i \ne {\cal I}$. In either case, $\mathbb{M}$ cannot be the first step of a perfect one-way LOCC protocol. 
\qed

\subsection{d = 2 + 3r}\label{Type II LOCC}
Let $d = 2 + 3r$ for $r \ge 1$. We will again use the $2 \times 2$ Pauli matrices $X,Y$ and $Z$ and fixed generic phases $\omega$ and $\gamma$ with $|\omega| = |\gamma| = 1$. We will also make use of the permutation matrix
\be
P = \begin{pmatrix}0&0&1 \\ 1& 0 & 0 \\ 0 & 1 & 0 \end{pmatrix}
\ee
and define the $3r \times 3r$ matrix $Q = P \ot {\cal I}_r$. We consider the set of matrices ${\cal I}_d,U,V$ with
\bee
U = \begin{pmatrix}\omega X & 0 \\ 0 &  Q \end{pmatrix} \qquad V = \begin{pmatrix} \gamma Z & 0 \\ 0 &  Q^2\end{pmatrix}
\eee
Let $\vert \psi_0 \ket$ be the standard maximally entangled state and choose $\omega$ and $\gamma$ with $ \gamma \ne \pm i\omega^2$. We claim that the states $\vert \psi_0 \ket$, $\vert \psi_1 \ket = (I \ot U) \vert \psi_0 \ket$ and $\vert \psi_2 \ket = (I \ot V) \vert \psi_0 \ket$  are orthogonal and maximally entangled but not distinguishable with one-way LOCC. In the case $d = 5$, these three states are given by
\be
\vert \psi_0 \ket &=& \frac{1}{\sqrt{5}} \left( |00\ket + |11\ket + |22\ket + |33\ket + |44\ket \right)_{AB} \nonumber \\
\vert \psi_1 \ket &=& \frac{1}{\sqrt{5}} \left( \omega |01\ket + \omega |10 \ket + |23\ket + |34\ket + |42\ket\right)_{AB} \label{d5 example} \\
\vert \psi_2 \ket &=& \frac{1}{\sqrt{5}} \left( \gamma |00\ket -\gamma |11\ket + |24\ket + |32\ket + |43\ket \right)_{AB} \nonumber
\ee
To show that one-way LOCC is insufficient, we will again suppose that Alice performs a measurement $\mathbb{M}$ on her system and receives the outcome corresponding to the operator $M^T$. If we write $M \ge 0$ in block form as $M = \begin{pmatrix}A & C^* \\ C & B \end{pmatrix}$, the fact that $\bra \psi_i \vert (I \ot M)\vert \psi_j \ket = 0$ for $i \ne j$ implies that 
\be
\tr MU &=& \omega \tr AX +  \tr BQ = 0\nonumber \\
\tr MV &=& \gamma \tr AZ +  \tr BQ^2 = 0\label{OddOrthConditions} \\
\tr MU^*V &=& -i \overline{\omega}\gamma \tr AY + \tr BQ = 0\nonumber
\ee
The first and third equations imply that $\omega \tr AX +i \overline{\omega}\gamma \tr AY= 0$. Since $\tr AX$ and $\tr AY$ are real, either $i\overline{\omega}^2\gamma$ is real or else $\tr AY = \tr AX = 0$. Since we assumed that $\gamma \ne \pm i\omega^2$,  $\tr AY = \tr AX = 0$, which implies that $\tr BQ = 0$, $\tr BQ^2 = \overline{\tr BQ} = 0$, and $\tr AZ = -\overline{\gamma}\tr BQ^2 = 0$ 

Hence, as in the previous example, we get that $A$ must be a multiple of the $2 \times 2$ identity matrix, which implies that either $M$ has rank greater than one or else $A = 0$. Again, we see that either $\mathbb{M}$ contains measurements of rank greater than one or else $\sum_i M_i \ne {\cal I}$. In either case, $\mathbb{M}$ cannot be the first step of a perfect one-way LOCC protocol.  \qed

\section{Positive Partial Transpose Measurements}\label{PPTSection}
Recall that the partial transpose of a matrix acting on a bipartite system is the application of the transpose map to just one of the two pieces of the system. While the transpose map ${\cal T}$ is positivity-preserving, the partial transpose map $(I \ot {\cal T})$ is not. On other hand, $(I \ot {\cal T})$ {\em is} positivity-preserving when restricted to {\em separable} matrices. Thus, if $M \in \mathbb{C}^d \ot \mathbb{C}^d$ is a separable operator, then $(I \ot {\cal T})M \ge 0$. Said differently, having a positive partial transpose is a necessary (but not sufficient) condition to being separable. We say that a measurement $\mathbb{M} = \{ M_1, M_2, \ldots, M_k\}$ is a positive partial transpose measurement (PPT) if $(I \ot {\cal T})M_i \ge 0$ for all $i$; this condition is necessary for $\mathbb{M}$ to be implemented using LOCC. 

For any pure maximally entangled state $\rho = \vert \psi \kb \psi \vert$, its partial transpose $\rho^{PT} = (I \ot {\cal T})\rho$ has eigenvalues $\frac{1}{d}$ and $-\frac{1}{d}$. Let $\{ \vert \psi_1 \ket, \vert \psi_2 \ket, \ldots \vert \psi_k \ket\}$ be an orthogonal set of pure maximally entangled bipartite states. We propose the following measurement to distinguish them:   \bee \mathbb{M} &=& \{ M_1, M_2, \ldots, M_k\} \\ M_i& =& \frac{1}{k}\left( {\cal I} + (k-1) \rho_i - \sum_{j \ne i} \rho_j \right) 
%\\ & = & \frac{1}{k} {\cal I} + \rho_i - \overline{\rho}
\eee
Since the $\rho_i$ are mutually orthogonal, $\sum_{j} \rho_j \le {\cal I}$ and $M_i \ge 0$.  It is clear that this is a POVM, since   $\sum_i M_i = {\cal I}$. 
Since all $\rho_j^{PT} \in [-\frac{1}{d}, \frac{1}{d}]$, we get  $\sum_{j \ne i} \rho_{j}^{PT} \le \frac{(k-1)}{d}$ and  \be
 (I \ot {\cal T}) M_i &\ge&  \frac{1}{k}\left( 1 -\frac{2(k-1)}{d} \right) \ge 0  \ee
whenver $\frac{2(k-1)}{d}\le 1$. 

This gives us our result: 
\begin{thm}\label{PPTTheorem}
Given $k$ orthogonal states $\{ \vert \psi_1 \ket, \vert \psi_2 \ket, \ldots, \vert \psi_{k} \ket\}$ which are maximally entangled on $\mathbb{C}^d \ot \mathbb{C}^d$. 

If $k \le \frac{d}{2} + 1$, then there exists a PPT measurement $\mathbb{M} = \{M_1, M_2, \ldots M_{k}\}$ such that 
\be
\bra \psi_i \vert M_j \vert  \psi_i \ket = \delta_{i,j}
\ee
\end{thm}
Specific to our current work, we set $k = 3$: 
\begin{cor}
Every set of three mutually orthogonal maximally entangled states in $\tensor[d]$ for $d \ge 3$ can be perfectly distinguished using a PPT measurement. 
\end{cor}
The corollary follows from the theorem when $d \ge 4$ and from the LOCC result in \cite{Nathanson2005} when $d = 3$. 

Note that Cosentino \cite{Cosentino} has given an example of $d = 2^n$ so-called ``lattice states'' in $\mathbb{C}^d \ot \mathbb{C}^d$ which {\it cannot} be distinguished with PPT measurements. This means that there is a least upper bound  $\alpha \in [\frac{1}{2},1]$ such that, for large enough $d$, any set of fewer than $\alpha d$ orthogonal maximally entangled states can be distinguished with PPT.

\section{Achieving the one-way  bound in Theorem \ref{LOCC1 Bound}}\label{OneWayConstruction}
In what follows we will assume that each $\vert \psi_i\ket$ occurs with a priori probability $p_i$, with $p_0 \ge p_1 \ge p_2$. To prove the bound, we use the naive strategy of perfectly distinguishing the more likely states $\vert \psi_0 \ket$ and $\vert \psi_1 \ket$ from each other and identifying the state as $\vert \psi_2 \ket$ only when our measurement is inconsistent with any other hypothesis. (This gives conclusive identification of $\vert \psi_2\ket$ as described in \cite{Bando3}.) By Theorem 2 in \cite{BN2013}, this procedure is guaranteed to give us a success probability of at least $p_0 + p_1 \ge \frac{2}{3}$, since we will be correct whenever the true state is either $\vert \psi_0 \ket$ and $\vert \psi_1 \ket$. However, we can do better. 

As above, we write $\vert \psi_i \ket = (I \ot U_i) \vert \psi_0 \ket$ for unitary matrices $U_i$ with $U_0 = I$. Without loss of generality, we assume that $U_1$ is diagonal in the standard basis. (Otherwise, Alice and Bob can perform a rotation of the form $(\overline{V} \ot V)$ to diagonalize $U_1$ and retain $U_0 =  I$.) As in \cite{Nathanson-2010}, we randomize the measurement by choosing a diagonal unitary $W_x = \sum_k e^{2\pi i x_k} \vert k \kb k \vert$, with $x = (x_0, x_1, \ldots, x_{d-1})$ chosen uniformly from $[0,1]^d$. Alice and Bob begin by performing the product rotation $W_x \ot W_x^*$, which leaves the states $\vert \psi_0\ket$ and  $\vert \psi_1 \ket$ invariant. 

Alice then performs a Von Neumann measurement in the basis $\{ \vert \varphi_j \ket \}_{j = 0,\ldots,d-1}$, where  \be \vert \varphi_j\ket =
\frac{1}{\sqrt{d}} \sum_{k = 0}^{d-1}e^{2\pi i jk /d} \vert k \ket. \ee 
If Alice gets the result $j$, then Bob measures using any basis which includes the orthogonal states $\vert \varphi_{d-j}\ket$ and $U\vert \varphi_{d-j}\ket$. This gives us the one-way LOCC measurement
\be
\Pi_0(x) &=& \sum_j W_x\vert \varphi_j \kb \varphi_j \vert W_x^* \ot  W_x^*\vert \varphi_{d-j} \kb \varphi_{d-j} \vert W_x  \nonumber \\
\Pi_1(x) &=& \sum_j W_x\vert \varphi_j \kb \varphi_j \vert W_x^*\ot  W_x^* U\vert \varphi_{d-j} \kb \varphi_{d-j} \vert U^* W_x \label{LOCC Projection}\\
\Pi_2(x) &=& {\cal I} - \Pi_0(x) - \Pi_1(x) \nonumber
\ee
This is a one-way LOCC measurement which distinguishes $\vert \psi_0\ket$ and $\vert \psi_1\ket$. If we average over all possible choices of $x$, these operators can be rewritten as 
\bee
\Pi_0 & = & \int_{[0,1]^d} \Pi_0(x) dx =  \vert \psi_0 \kb \psi_0 \vert + \frac{1}{d} R \\
\Pi_1 &=& \int_{[0,1]^d} \Pi_1(x) dx =  \vert \psi_1 \kb \psi_1 \vert + \frac{1}{d} R
\eee
where $R = \sum_{i \ne j} \vert i \ot j \kb i \ot j \vert$. This makes it clear that $\bra \psi_0 \vert \Pi_0 \vert \psi_0 \ket = \bra \psi_1 \vert \Pi_1 \vert \psi_1 \ket = 1$. We also see that $\bra \psi_2\vert \Pi_0 \vert \psi_2 \ket=\bra \psi_2\vert \Pi_1 \vert \psi_2 \ket  = \frac{1}{d} \bra \psi_2 \vert R \vert \psi_2 \ket \le \frac{1}{d} || R ||_\infty = \frac{1}{d}$. Hence, the probability of error is bounded by
\be
P_{error} =  p_2 \bra \psi_2\vert (\Pi_0 + \Pi_1)\vert \psi_2 \ket \le \frac{2}{3d} \label{errorbound}
\ee
\qed

\section{More than three states} \label{MoreStates}
We would like to show that the phenomenon described here is not limited to three states. That is, for fixed $k \ge 3$, we wish to find a set of $k$ orthogonal maximally entangled states in high dimension which cannot be distinguished using one-way LOCC. This can always be done. 
\begin{prop}
For any $k \ge 3$ and for arbitrarily large values of $d$, there exist sets of $k$ orthogonal maximally entangled states in $\tensor[d]$ which are perfectly distinguishable with PPT measurements but not with one-way LOCC.
\end{prop}
We build such an example below, which is a generalization of the one in Section \ref{Type II LOCC}, and show that it cannot be distinguished with one-way LOCC. Note that if $d$ is large enough, then Theorem \ref{PPTTheorem} guarantees that these states can be distinguished with PPT measurements. 

Let $m = 2^n$ be a power of two with $k \le m^2$; let $\{\vert \phi_i \ket = (I \ot X_i)\vert \Phi_m \ket \}$ be a subset of the qubit lattice states which cannot be distinguished using one-way LOCC. Any set of $k >m$ lattice states is locally indistinguishable; examples with $k = m$ are given in \cite{Cosentino, FourQuquad}; and an example with $k=15<d = 16$ is given in \cite{SmallSets}. Sets of lattice states have the property that each $X_i$ is a hermitian matrix and each product $X_iX_j$ is either hermitian or skew hermitian. We wish to show that these states can be used to build sets of $k$ states in arbitrarily large dimensions which cannot be distinguished with one-way LOCC.  

As in our earlier example with $k = 3$, define $P = \sum_{j = 0}^{k-1} \vert j+1 \kb j \vert$ (with addition modulo $k$) and $Q = P \ot {\cal I}_r$ for $r \ge 1$. Then our set of states is $\{ \vert \psi_i \ket = (I \ot U_i)\vert \Phi_{d} \ket \}_{i = 0, \ldots, k-1}$ with $d = m + kr$ and 
\bee
U_i = \begin{pmatrix} \alpha_i X_i & \\ & Q^i \end{pmatrix} 
\eee
where the partition is into blocks of size $m$ and $kr$ and the $\alpha_i$ are arbitrary points on the unit circle. Since $\tr U_i^*U_j = \tr \left(\alpha_j\overline{\alpha_i} X_iX_j + Q^{j-i} \right)= kr\delta_{i,j}$, these states are orthogonal.  We will show that if there exists a one-way LOCC measurement to distinguish the $\{ \vert \psi_i \ket \}$, then there exists one to distinguish the  $\{ \vert \phi_i \ket \}$, contradicting our assumption. 

The proof follows the one in Section \ref{Type II LOCC}. If $\mathbb{M}$ is an initial measurement performed by Alice, with $M = \begin{pmatrix}A & C^* \\ C & B\end{pmatrix} \in \mathbb{M}$, then for all $1 \le j < k-1$
\bee
 \tr U_0MU_j^* &=& \alpha_0\overline{\alpha_j} \tr AX_0X_j + \tr BQ^{-j} =0   
 \\ \tr U_1MU_{j+1}^*& =& \alpha_1\overline{\alpha_{j+1}} \tr AX_1X_{j+1} + \tr BQ^{-j} =0
\eee
This means that 
\bee
 \alpha_0\overline{\alpha_j} \tr AX_0X_j  = \alpha_1\overline{\alpha_{j+1}} \tr AX_1X_{j+1}
 \eee
We know that $\tr AX_0X_j$ and $\tr AX_1X_{j+1}$ are either real or pure imaginary, which implies that $(\alpha_0\overline{\alpha_j}\alpha_{1}\overline{\alpha_{j+1}} )^4=1$  or else $\tr AX_0X_j  =0$. Since the $\alpha_i$ are chosen generically, we conclude that $\tr AX_0X_j  =\tr AX_1X_{j+1}= 0$. This means that $\tr BQ^j =0$ for all $1 \le j \le k-1$ and hence $\tr AX_iX_{j}= -\overline{\alpha}_i\alpha_j \tr BQ^{i-j} = 0$ for all $i \ne j$. 

So, if there exists a one-way LOCC measurement which distinguishes the $\vert \psi_i \ket$, then Alice's measurement $\mathbb{M} = \{M_k\}$ consists of rank one matrices such that the corresponding upper components $A_k$ satisfy $\tr A_kX_iX_{j} = 0$ when $i \ne j$. Such $A_k$ must have rank at most one and satisfy $\sum_{k} A_k = {\cal I}_m$; but this implies that $\mathbb{A} = \{A_i\}$ is the first step in a one-way LOCC measurement which perfectly distinguishes the $\vert \phi_i \ket$. 

Since we assumed that no such measurement exists for the $\{\vert \phi_i \ket \}$, none exists for the $\{ \vert \psi_i \ket\}$ and we have a set of $k$ maximally-entangled states in $\tensor[m+ kr]$ for which perfect one-way discrimination is impossible.  And since we can make $r$ as large as we like, we can have $k$ maximally entangled states in arbitrarily large spaces (and hence with arbitrarily large amounts of entanglement) which cannot be distinguished with one-way LOCC.  
\qed

It is hoped that specific constructions of such sets of states will establish (a) whether these states are distinguishable with two-way LOCC and (b) whether these sets are not simply supersets of each other; that is, can we find such sets for which any $\frac{k}{2}$ of them {\it are} distinguishable with one-way LOCC. While this has not been done in general, we can say something stronger in the case $k = 4$. 
\begin{prop}
In arbitrarily high dimensions $d$, there exists sets of four maximally entangled states such that any three of them can be distinguished with one-way LOCC but the entire set cannot be. \end{prop}
We use the construction above with $k = m = 4$ and use the specific example from \cite{FourQuquad, Cosentino} of a set of four lattice states which are not distinguishable even with PPT measurements. The rest of the proposition follows from the following lemma, which is proved in the appendix. 
\begin{lemma}\label{Lattice}
Any set of three qubit lattice states in $\tensor[4]$ can be distinguished with one-way LOCC.
\end{lemma}
This means that the set of four indistinguishable lattice states in $\tensor[4]$ is minimal, since any three can be distinguished with one-way LOCC.
 
\section{Conclusions} 
We have addressed the question of whether there exist triples of orthogonal maximally entangled states in $\tensor[d]$ which cannot be distinguished using local operations and classical communications; and in the process have given the smallest possible examples of sets of states which are distinguishable with two-way LOCC but not one-way. The region between these two paradigms remains elusive, and very few examples have been given of sets such as these. On the other hand, we have shown that {\it any} set of three orthogonal maximally entangled states can be distinguished perfectly with a positive partial transpose measurement and with high probability using one-way LOCC. 

The answer to the question of LOCC distinguishability is located in the small space between these results, somewhere in the murky area between LOCC-1 and PPT. The open question of the existence of triples which cannot be distinguished with LOCC is interesting for its own sake but also to push our understanding of the line between LOCC and not. 

\medskip

\noindent \textbf{Acknowledgement:} The author is grateful
to Saint Mary's College for granting a sabbatical, during which much of this
work was completed.

\appendix
\section{Distinguishing the examples using two-way LOCC}
Given the difficulty in giving a mathematical characterization of the full set of LOCC measurements, the most effective way to show that a set of states is distinguishable with LOCC is to explicitly build a protocol to distinguish them. Below, we show how we can distinguish the examples described in Section \ref{CounterExamples}. 

\subsection{A two-way LOCC protocol to distinguish the  $d =2m$ example} \label{TwoWayEven}
We show that the examples given in Section \ref{CounterexamplesEven} may be distinguished perfectly with a two-way LOCC protocol. We use the tensor product structure of this example to use one-way LOCC to eliminate one of the three states; that is, we use one-way LOCC to transform $\{ \vert \psi_0 \ket, \vert\psi_1 \ket, \vert \psi_2 \ket\}$ into $\{ \vert \Phi_{j} \ket, \vert \Phi_{k} \ket\}$ for $\{ j,k \} \subset \{0,1,3\}$. We can then use a second round of LOCC to distinguish the remaining two options. The protocol is as follows: 

(1) We will initially assume that the origin is in the convex hull of the points $\{1, \omega, \gamma\}$ in the complex plane. This is equivalent to the fact that the imaginary parts of $\overline{\omega}, \gamma$ and $\overline{\gamma}\omega$ are either all positive or all negative.  

Suppose this is the case. Then we can define
\be\label{pvector}
p_0 &=& \left\vert \Im( \overline{\gamma}\omega) \right\vert \qquad
p_1 = \left\vert \Im(\gamma)\right\vert \qquad
p_2 = \left\vert \Im(\omega) \right\vert 
\ee
with $p = (p_0,p_1, p_2) \in \mathbb{R}_+^3$.  It is easily checked that under our assumption,  $p$ is orthogonal to the vector $(1,\omega, \gamma)$. 

(2)  Alice performs the measurement $\mathbb{A} = \{A_{j,k}^T\}_{j = 0 \ldots m-1, k = 0,1}$ which acts only on her first qubit system, defined by
\bee
A_{0,0} &=& \frac{m-2}{m-1}\left( {\cal I}_m - \vert 0 \kb 0 \vert \right) \ot {\cal I}_2 \\
A_{j,k} & = & \frac{1}{m-1} \vert a_{j,k} \kb a_{j,k} \vert \ot  {\cal I}_2 \qquad \mbox{ for } j>0
\\ &&\vert a_{j,k} \ket= \frac{1}{\sqrt{2}} \left( \vert 0 \ket + (-1)^k \vert j\ket \right)
\eee
If Alice receives the outcome $A_{0,0}$, then the first system is in the same entangled state $\frac{1}{\sqrt{m-1}} \sum_{j}\vert jj\ket_{A_1B_1}$ regardless of the value of $\vert \psi_i\ket$. Since $m>2$, we can use this state to teleport half of the Bell state from Alice to Bob, and Bob can distinguish the three Bell states once they are completely on his side. In this case, only one direction of LOCC is needed. 

However, suppose Alice receives the outcome $A_{j,k}$ for $j>0$. Then Bob's first qubit system is in the state $T_\alpha \vert a_{j,k} \ket$ for $\alpha \in \{1,\omega, \gamma\}$, which lies in the two-dimensional span of $\vert 0 \ket$ and $\vert j \ket$. Bob then measures this subspace using the three-outcome POVM
\bee
\mathbb{B}_{j,k} &=&  \frac{2}{p_0+ p_1 + p_2} \{ p_i \vert b_{j,k,i} \kb b_{j,k,i}|  \ot {\cal I}_2\}_{i=0,1,2} \\
\vert b_{j,k,0} \ket &=& \frac{1}{\sqrt{2}} \left( \vert 0 \ket + (-1)^k \vert j\ket \right)\\
\vert b_{j,k,1} \ket &=& \frac{1}{\sqrt{2}} \left( \vert 0 \ket + (-1)^k \overline{\omega} \vert j\ket \right)\\
\vert b_{j,k,2} \ket &=& \frac{1}{\sqrt{2}} \left( \vert 0 \ket + (-1)^k \overline{\gamma}\vert j\ket \right) %\{ p_0\pmatrix{ 1 & -1 \cr -1 & 1}, p_1\pmatrix{ 1 & -\gamma \cr -\overline{\gamma} & 1}, p_2\pmatrix{ 1 & -\omega \cr - \overline{\omega} & 1}\}
\eee
The sum of the off-diagonal terms is given by \bee \frac{1}{p_0+p_1 + p_2}\sum_i (-1)^k(p_0 + p_1 \overline{\omega} + p_2\overline{\gamma}) = 0\eee by our construction of the vector $p$ in (\ref{pvector}). This means that \bee \sum_i 
\frac{2p_i }{p_0+ p_1 + p_2} \vert b_{j,k,i} \kb b_{j,k,i} \vert = \vert 0 \kb 0 \vert + \vert j \kb j \vert\eee
So,  $\mathbb{B}_{j,k}$ is a complete measurement on this two-dimensional space; and for each $i$, $\bra a_{j,k} \ot b_{j,k,i} \vert \psi_i \ket = 0$. So, no matter the outcome, one of the possibilities for $i$ has been eliminated. 

(3) Now, Alice and Bob can dispose of their $m\times m$ system and focus on their qubit system, where they share one of two possible Bell states, unchanged by the previous measurements. These two Bell states can be perfectly distinguished with LOCC. 
 
Note that Bob's measurement depends on some information from Alice and also that Alice's measurement depends on information from Bob. This is the heart of a two-way LOCC protocol. 

(4) We now return to our initial assumption, that the origin was in the convex hull of $\{1, \omega, \gamma\}$. Without this assumption, the nonnegative vector $(p_0, p_1, p_2)$ is not orthogonal to $(1, \omega, \gamma)$. 

To fix this, we perform a rotation on each of Alice and Bob's systems. This is the only step in which the $m \times m$ systems interact with the qubit ones. For each $\sigma_j \in \{{\cal I}_2, X,Y,Z\}$, we define a block diagonal unitary matrix:
\bee
W_j = \begin{pmatrix} \sigma_j & \\ & {\cal I}_{m-1} \ot {\cal I}_2\end{pmatrix} \eee
That is, all the diagonal blocks are the identity except the first. This is essentially a control-$\sigma_j$ of a $m$-dimensional system on a two-dimensional system: If the large system is in state $\vert 0 \ket$, $\sigma_j$ is applied to the small system. We note that for any phase $\alpha$, $W_j (T_\alpha \ot \sigma_k) W_j^* = T_{\pm \alpha} \ot \sigma_k$.  

So, as an initializing first step, Alice and Bob perform the product rotation $\overline{W_j} \ot W_j$, which affects the states as 
 \bee
\{ T_1 \ot {\cal I}_2, T_\omega\ot X, T_\gamma\ot Z\} \rightarrow \{ T_1 \ot {\cal I}_2, T_{ \omega'}\ot X, T_{\gamma'} \ot Z\} 
\eee
with $\omega' = \pm \omega, \gamma' = \pm \gamma$. We choose the unique value of $j \in \{0,1,2,3\}$ such that the imaginary parts of $\overline{\omega'}, \gamma'$ and $\overline{\gamma'}\omega'$ are either all positive or all negative. After this step, the origin is in the convex hull of $\{1, \omega', \gamma'\}$ and we can run the rest of our protocol.  

We can see then that the three states $\{\vert\psi_i \ket\}$ may be perfectly distinguished with two-way LOCC, even though it is not possible with one-way. \qed

\subsection{Two-way LOCC protocol to distinguish the examples in Section \ref{Type II LOCC} }
We now show that the examples in Section \ref{Type II LOCC} can also be distinguished with two-way LOCC. Unlike the previous protocol, this is not a method of elimination: All three possibilities remain until the last step. The protocol is as follows. 

First, Alice measures using $\mathbb{A} = \{A_k^T\}_{k =0, \ldots 3r-1} $ with \bee A_k = \frac{1}{3r}\vert 0 \kb 0 \vert + \frac{1}{3r}\vert 1 \kb 1 \vert + \vert k+2 \kb k+2 \vert\eee All outcomes are equivalent, so without loss of generality we will assume that $k = 0$. And for simplicity, we will focus on the base case, $r = 1$; the general case is identical except that all threes are replaced with $3r$. The outcome $A_0$ maps the matrix $U_i$ to $U_i A_0^{1/2}$, leaving our three matrices as 
\bee
\begin{pmatrix} \frac{1}{\sqrt{3}} {\cal I}  & \\ & \vert 0 \kb 0 \vert\end{pmatrix}\qquad \begin{pmatrix} \frac{1}{\sqrt{3}} \omega X & \\ & \vert 1 \kb 0 \vert\end{pmatrix} \qquad \begin{pmatrix} \frac{1}{\sqrt{3}} \gamma Z & \\ & \vert 2 \kb 0 \vert \end{pmatrix}
\eee
We claim that at this point, only one more measurement by each party is necessary. According to Proposition \ref{OneWayLOCC}, this is equivalent to the existence of a partial isometry $W_0$ such that for each $i \ne j$, $W_0U_iA_kU_j^*W_0^*$ has every diagonal element equal to zero. We construct such a $W$ below, with some motivation. 

Let  $(u,  v)^*$ be a row of $W$, partitioned with $u \in \mathbb{C}^2$ and $v \in \mathbb{C}^3$ as before. The orthogonality conditions imply that 
\be \label{TwoWay5Orthogonality}
\omega u^*Xu + 3 \bra 0 \vert vv^* \vert 1 \ket = 0 \nonumber \\
\gamma u^*Zu + 3 \bra 0 \vert vv^* \vert 2 \ket = 0\\
\omega \overline{\gamma} u^*ZXu + 3 \bra 2 \vert vv^* \vert 1 \ket = 0 \nonumber
\ee
We write $v = \begin{pmatrix}v_0 \\ v_1 \cr v_2\end{pmatrix}$ and mimic the argument in Section \ref{Type II LOCC}, noting that (\ref{TwoWay5Orthogonality}) implies that
\bee
\{ \overline{\omega} \overline{v_0}v_1 , \overline{\gamma} \overline{v_0}v_2, i\overline{\omega} \gamma \overline{v_2}v_1\} \subset \mathbb{R}
\eee
For general $\omega$ and $\gamma$, this implies that two of these numbers must be zero. Hence, $v_k = 0$ for some $k \in \{ 0,1,2\}$. This also implies that $u^*\sigma_i u = 0$ for two values of $i$. 

As an example, suppose that $v_1 =0$. This implies that $u^*Xu = u^*Yu = 0$, which means that either $u  =0$ or else $u$ is an eigenvector of $Z$.  If $u \ne 0$, then (\ref{TwoWay5Orthogonality}) implies
\bee
0 = \gamma u^*Zu + 3\overline{v_0}v_2= \pm \gamma u^*u+ 3 \overline{v_0}v_2\eee
So $\overline{v_0}v_2 = \mp \frac{\gamma}{3} u^*u$. This leads us to the following set of 4 (non-normalized) vectors: 
\bee
\begin{pmatrix}u_i \\ v_i \end{pmatrix} \in \left\{  \begin{pmatrix} \sqrt{3} \\ 0 \\ 1 \\ 0 \\ -\overline{\gamma}\end{pmatrix},\begin{pmatrix} -\sqrt{3} \\ 0 \\ 1 \\ 0 \\ -\overline{\gamma}\end{pmatrix},\begin{pmatrix} 0 \\ \sqrt{3} \\ 1 \\ 0 \\ \overline{\gamma}\end{pmatrix},\begin{pmatrix} 0 \\ -\sqrt{3}  \\ 1 \\ 0 \\ \overline{\gamma}\end{pmatrix}  \right\}
\eee
Each of these vectors has the desired properties. In addition, 
\bee
\sum_{i = 0}^3 \begin{pmatrix}u_i \\ v_i\end{pmatrix}\begin{pmatrix} u_i^* & v_i^*\end{pmatrix} = \begin{pmatrix}6 &&&& \\ & 6 &&& \\ && 4 && \\ &&& 0 & \\ &&&& 4\end{pmatrix}
\eee
We can similarly define sets of four vectors each based on the eigenvectors of $X$ and of $Y$. This gives us twelve vectors; summing up all of them gives
\bee
\sum_{i = 0}^{11} \begin{pmatrix}u_i \\ v_i\end{pmatrix}\begin{pmatrix}u_i^* & v_i^*\end{pmatrix} = \begin{pmatrix}18 &&&& \\ & 18 &&& \\ && 8 && \\ &&& 8 & \\ &&&& 8\end{pmatrix}
\eee
To complete the measurement, we allow the case $u = 0$, in which case $v = \sqrt{10} \vert k \ket$ is simply a multiple of a standard basis vector. Rescaling gives us a complete measurement with 15 outcomes which Bob can perform based on the outcome of Alice's measurement. (If Bob didn't know the outcome of Alice's measurement, he couldn't properly coordinate the values of $u$ and $v$, hence the necessitity of two-way communication.) The corresponding partial isometry has four columns for each of the Pauli matrices plus a set of three columns for the identity, yielding the $5 \times 15$ matrix $W_0^*$: 
 \bee 
\frac{1}{\sqrt{18}} \begin{pmatrix}\sqrt{3} & - \sqrt{3} & 0 & 0 & \sqrt{\frac{3}{2}}  &- \sqrt{\frac{3}{2}} & \sqrt{\frac{3}{2}} & -\sqrt{\frac{3}{2}} & \sqrt{\frac{3}{2}} &- \sqrt{\frac{3}{2}} & \sqrt{\frac{3}{2}} & -\sqrt{\frac{3}{2}}&0&0&0 \\ 0 & 0 & \sqrt{3} & - \sqrt{3} & \sqrt{\frac{3}{2}}  &- \sqrt{\frac{3}{2}} & -\sqrt{\frac{3}{2}} & \sqrt{\frac{3}{2}} & -i\sqrt{\frac{3}{2}} &i \sqrt{\frac{3}{2}} & i\sqrt{\frac{3}{2}} & -i \sqrt{\frac{3}{2}}&0&0&0 \\ 1&1&1&1&1&1&1&1&0&0&0&0&\sqrt{10}&0&0 \\ 0&0&0&0&- \overline{\omega}&- \overline{\omega}& \overline{\omega} & \overline{\omega}
& 1 & 1& 1 & 1& 0 & \sqrt{10} &0 \\ -\overline{\gamma} &-\overline{\gamma} &\overline{\gamma} &\overline{\gamma} & 0 & 0 & 0 & 0 & - i \omega\overline{\gamma} &- i \omega\overline{\gamma} & i \omega\overline{\gamma} & i \omega\overline{\gamma} & 0 & 0 & \sqrt{10}\end{pmatrix}
 \eee
It is straightforward to check that $W_0^*W_0 = {\cal I}_5$ and that when $i \ne j$, the diagonal entries of $W_0U_iA_0U_j^*W_0^*$ are all zero. This means that Bob can apply the partial isometry to embed his 5-dimensional system in a 15-dimensional one. He then measures in the standard basis and gets the outcome $x$. Alice's task now is to distinguish the states $A_0^{1/2} U_i^T W_0^T \vert x \ket$, which are mutually orthogonal.

\section{Proof of Lemma \ref{Lattice}}
Suppose our three lattice states are given by 
\bee
\vert \varphi_i \ket = \vert \Phi_{x(i)} \ket \ot  \vert \Phi_{y(i)} \ket
\eee
with $i = 0,1,2$ and $x(i), y(i) \in \{0,1,2,3\}$. There are two possibilities: 

Suppose $x(0) = x(1) = x(2)$. Then we can use the first system to teleport the second to Bob, who can perform a complete local measurement to distinguish  the $\{ \vert \Phi_{y(i)} \ket \}$, which must all be distinct. Alice's only measurements happen in the teleportation, so this is accomplished with one-way LOCC. Likewise, if $y(0) = y(1) = y(2)$. 

If this is not the case, then there must be a relabeling of the states $\{0,1,2\}$ such that $x(1) \notin \{x(0), x(2) \}$ and $y(2) \notin \{y(0), y(1) \}$. These are just Pauli states, and there exist one-way LOCC measurements to distinguish $\vert \Phi_{x(1)} \ket$ from the other two states $\{ \vert \Phi_{x(0)} \ket, \vert \Phi_{x(2)}\ket \}$ on the first system; and to distinguish $\vert \Phi_{y(2)} \ket$ from the other two states $\{ \vert \Phi_{y(0)} \ket, \vert \Phi_{y(1)}\ket \}$ on the second. Alice performs the first step of these two measurements in parallel, reports her outcomes to Bob, who can then complete each measurement. This uniquely identifies $\vert \varphi_i \ket$, as seen in the table below: 

\bee \begin{array}{c|c|c| } & x(0), x(2)  & x(1) \cr \hline y(0),y(1) & \vert \Phi_0 \ket&\vert \Phi_1 \ket \cr \hline y(2)  &\vert \Phi_2 \ket & \cr \hline
\end{array}\eee

Regardless of the outcome, we can distinguish three lattice states in $\tensor[4]$ using one-way LOCC.


\begin{thebibliography}{1}


\bibitem{data-hiding} D. P. DiVincenzo, D. W. Leung
and B. M. Terhal, 
    Quantum data hiding, IEEE Trans. Inf. Theory,
vol.48(3), pp. 580598 (2002). %   arXiv:quant-ph/0103098.

\bibitem{MatthewsWehnerWinter} W. Matthews, S. Wehner, A. Winter,
Distinguishability of quantum states under restricted families of
measurements with an application to quantum data hiding, Comm. Math.
Phys. \textbf{291}, Number 3 (2009).

\bibitem{Virmani-2001} S. Virmani, M. F. Sacchi, M. B. Plenio, D.
Markham, Optimal local discrimination of two multipartite pure states,
Phys. Lett. A. \textbf{288}  (2001). 

\bibitem{Walgate} J. Walgate, A. J. Short, L. Hardy, and V.
Vedral, Local distinguishability of multipartite orthogonal quantum
states, Phys. Rev. Lett. \textbf{85}, 4972 (2000).


\bibitem{Nathanson2005} M. Nathanson, Distinguishing bipartite
orthogonal states by LOCC: best and worst cases, Journal of Mathematical
Physics \textbf{46}, 062103 (2005).



\bibitem{Fan} H. Fan, Distinguishability and indistinguishability
by local operations and classical communication, Phys. Rev. Lett.
\textbf{92}, 177905 (2004).


\bibitem{Walgate-2002} J. Walgate, L. Hardy, ``Nonlocality, Asymmetry, and Distinguishing Bipartite States,'' Phys. Rev. Lett. \textbf{89}, 147901 (2002).

\bibitem{BGK} S. Bandyopadhyay, S. Ghosh, G. Kar, LOCC distinguishability of unilaterally transformable quantum states, New J. 
Phys. {\bf 13} 123013 (2011). %\textit{quant-ph/1102.0841}

\bibitem{Cosentino} A. Cosentino, Positive partial transpose
indistinguishable states via semidefinite programming, Phys. Rev.
A \textbf{87}, 012321 (2013).% \textit{quant-ph/1205.1031} 


\bibitem{FourQuquad} N. Yu, R. Duan and M. Ying  Four Locally Indistinguishable Ququad-Ququad Orthogonal Maximally Entangled States, Phys. Rev. Lett. \textbf{109}, 020506 (2012). %\textit{quant-ph/1107.3224}

\bibitem{OH-2006} M. Owari, M. Hayashi, ``Local copying and local discrimination
as a study for non-locality of a set,'' Physical Review A \textbf{74},
032108 (2006); Physical Review A \textbf{ 77}, 039901(E) (2008).


\bibitem{Ghosh-2001} S. Ghosh, G. Kar, A. Roy, A. Sen (De), and U.
Sen, Distinguishability of Bell states, Phys. Rev. Lett. \textbf{87},
277902 (2001).


\bibitem{Hayashi Bounds 2006} M. Hayashi, D. Markham, M. Murao, M. Owari, S. Virmani,
``Bounds on Multipartite Entangled Orthogonal State Discrimination
Using Local Operations and Classical Communication,''Physical Review
Letters, \textbf{ 96} 040501, (2006).


\bibitem{Bennett-I-99} C. H. Bennett, D. P. DiVincenzo, C. A. Fuchs,
T. Mor, E. Rains, P. W. Shor, J. A. Smolin, and W. K. Wootters, Quantum
Nonlocality without Entanglement, Phys. Rev. A \textbf{59}, 1070
(1999).

\bibitem{OwariHayashi} M. Owari, M. Hayashi, Two-way classical communication remarkably
improves local distinguishability, New J. Phys. {\bf 10} 013006 (2008). 
\bibitem{Bando3} S. Bandyopadhyay, J. Walgate, Local distinguishability of any three quantum states,
Journal of Physics A {\bf 42}, 072002 (2009). 


\bibitem{BN2013} S. Bandyopadhyay, M. Nathanson, Tight bounds on the distinguishability of quantum states under separable measurements, Phys. Rev. A {\bf 88}, 052313 (2013).


\bibitem{Nathanson-2010} M. Nathanson, Testing for a pure state with
local operations and classical communication, Journal of Mathematical
Physics \textbf{51} (2010) 042102.



\bibitem{SmallSets} A. Cosentino, V. Russo, Small sets of locally indistinguishable orthogonal maximally entangled states, \href{http://lanl.arxiv.org/abs/1307.3232}{\textit{quant-ph/1307.3232}}.













\end{thebibliography}
\end{document}